\documentclass[a4paper,11pt]{article}

\usepackage{jcappub} 

\usepackage[T1]{fontenc} 

\title{\boldmath Natural Inflation with a negative cosmological constant}

\author[a,1]{Chia-Min Lin,\note{Corresponding author.}}
\author[b,c]{Naoto Maki}
\author[c,d,b,e,f]{Kazunori Kohri}

\affiliation[a]{Fundamental General Education Center, National Chin-Yi University of Technology, Taichung 411030, Taiwan, R.O.C.}
\affiliation[b]{Department of Astronomical Science, The Graduate University for Advanced Studies (SOKENDAI), 2-21-1 Osawa, Mitaka, Tokyo 181-8588, Japan}
\affiliation[c]{Division of Science, National Astronomical Observatory of Japan, 2-21-1 Osawa, Mitaka, Tokyo 181-8588, Japan}
\affiliation[d]{Department of Astronomy, The University of Tokyo, Bunkyo-ku, Hongo, Tokyo 113-0033, Japan}
\affiliation[e]{Theory Center, IPNS, KEK,
1-1 Oho, Tsukuba, Ibaraki 305-0801, Japan}
\affiliation[f]{Kavli IPMU (WPI), UTIAS, The University of Tokyo, Kashiwa, Chiba 277-8583, Japan}
\emailAdd{cmlin@ncut.edu.tw}
\emailAdd{naoto.maki@grad.nao.ac.jp}
\emailAdd{kazunori.kohri@nao.ac.jp}

\abstract{In this work, we investigate a cosmic inflation model based on a cosine-type potential with a negative cosmological constant. This model originates from a classical solution of the Wheeler-DeWitt equation. The equation of motion for the inflaton field can be solved analytically without relying on approximation schemes, such as the slow-roll conditions. The predictions of the spectral index, the tensor-to-scalar ratio, and the running spectral index are calculated and compared with experimental constraints from Planck Collaboration, Atacama Cosmology Telescope Collaboration (ACT), and Dark Energy Spectroscopic Instrument (DESI).}

\begin{document}
\maketitle
\flushbottom

\section{Introduction}
\label{sec:Introduction}

Cosmic Inflation \cite{Starobinsky:1980te, Guth:1980zm, Linde:1981mu} is widely regarded as the standard paradigm for the very early universe. A period of accelerated expansion can resolve many problems of the conventional hot big bang model, such as the flatness problem, the horizon problem, the monopole problem, the gravitino problem, and other unwanted relic problems. Quantum fluctuations during inflation generate primordial density perturbations, which provide the seeds for structure formation. These perturbations also account for, and are constrained by, the anisotropies of the cosmic microwave background (CMB). At present, inflation should be regarded as a standard paradigm rather than a standard model, since a large number of inflationary models have been proposed.

Constructing a single-field inflation model with an inflaton field $\phi$ typically amounts to choosing a particular form of the inflaton potential $V(\phi)$, which is dictated by an underlying symmetry. In the framework of Einstein gravity, the equation of motion is given by
\begin{equation}
\ddot{\phi}+3H\dot{\phi}+V^\prime(\phi)=0,
\end{equation}
where a dot denotes differentiation with respect to time, and $H$ is the Hubble parameter, which is related to the inflaton field through the Friedmann equation in a flat universe,
\begin{equation}
\rho=3H^2=\frac{1}{2}\dot{\phi}^2+V(\phi).
\end{equation} 
Here and throughout the following discussion, we set the reduced Planck mass to unity, $M_P=1$. Depending on $V(\phi)$, the equation of motion is generally a complicated nonlinear differential equation. To simplify the analysis, one often introduces the slow-roll conditions,
\begin{equation}
3H\dot{\phi}+V^\prime(\phi) \simeq 0,
\end{equation}
and
\begin{equation}
3H^2 \simeq  V(\phi).
\end{equation}
Furthermore, slow-roll parameters such as $\epsilon$ and $\eta$ can be defined, allowing inflationary predictions, including the spectral index, to be expressed in terms of these parameters. In this work, however, we propose an inflationary model that does not require this conventional slow-roll analysis.

In our previous work \cite{Maki:2026fgm}, we demonstrated that imposing the condition $|\Psi|=1$ on the wave function of the universe in the Wheeler--DeWitt equation for a flat FRW minisuperspace, with matter described by a scalar field and its potential, restricts the allowed forms of the scalar potential to only three classes: the quadratic potential with a negative cosmological constant, the cosine-type potential with a negative cosmological constant, and the exponential potential (without a cosmological constant) or hyperbolic cosine potential (with either a positive or negative cosmological constant). 

The solution corresponding to the quadratic potential with a negative cosmological constant was first discovered in \cite{Lin:2023sza} and later applied to an inflationary scenario known as uniform rate inflation in \cite{Lin:2023xgs, Lin:2023ceu}. In the present work, we construct an inflation model based on the cosine-type potential with a negative cosmological constant. This kind of periodic potential may originate from an underlying symmetry, such as a $U(N)$ or $SU(N)$ symmetry. It is interesting to note, however, that no such symmetry was imposed explicitly in \cite{Maki:2026fgm}. This observation suggests the conditions imposed there may effectively give rise to such a symmetry in an implicit manner. Our model differs from Natural Inflation \cite{Freese:1990rb} because of the presence of the negative cosmological constant. We note that similar cosine-type potentials with a negative cosmological constant have been investigated in the context of dark energy \cite{Luu:2025fgw,Luu:2025dax,Shiu:2026edl,Murai:2025msx}.
At first sight, it may not appear significantly different from the original natural inflation model; however, the negative cosmological constant makes it possible to obtain analytical solutions for both the equation of motion and the time evolution of the scale factor. 

\section{Natural inflation with a negative cosmological constant}
\label{}
We consider a single field inflation model in which the scalar potential $V(\phi)$ is given by
\begin{align}
    V(\phi)&=A^2 \left\{ \left( \frac{3}{8}-\frac{\omega^2}{4} \right) - \left( \frac{3}{8}+\frac{\omega^2}{4} \right) \cos (2\omega \phi+2B) \right\},
\label{potential}
\end{align}
where $\omega =\sqrt{6\left( q-\frac{1}{4} \right)}$ with $q>\frac{1}{4}$, and $A$ and $B$ are arbitrary constants. 
This peculiar form\footnote{After completing this work, we noticed that the potential coincides with one of the potentials considered in constant-roll inflation \cite{Motohashi:2014ppa}, although it is derived from a different ansatz.} arises from an exact solution of the Wheeler–DeWitt Equation in a flat FLRW minisuperspace under the condition that the wave function of the universe satisfies $|\Psi|=1$. This condition is so restrictive that only three classes of potentials are allowed \cite{Maki:2026fgm}. As mentioned in the Introduction, the condition may effectively give rise to an underlying symmetry in an implicit manner, thereby leading to a periodic potential. The parameter $q$ is a real constant for ambiguity in operator ordering.

\begin{figure}[t]
  \centering
\includegraphics[width=0.6\textwidth]{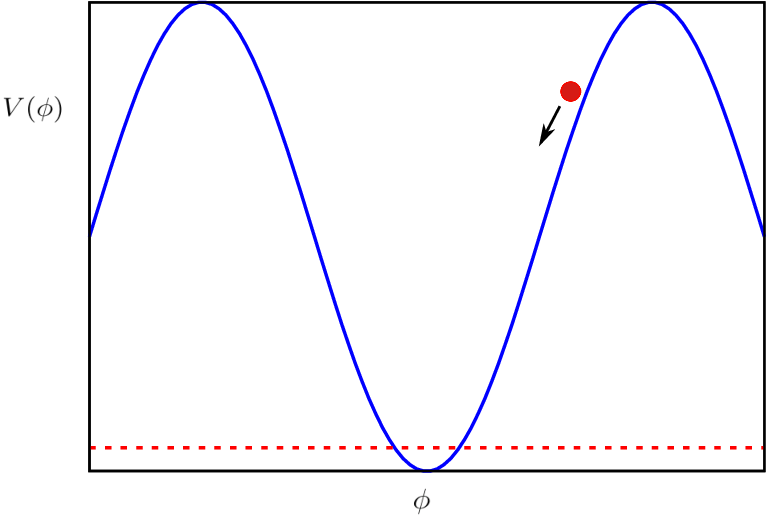}
  \caption{The potential shape.}
  \label{po}
\end{figure}

In the following discussion, we set $B=0$, corresponding to the choice that $\phi=0$ is located at the minimum of the potential. In this case, the vacuum corresponds to an anti-de Sitter vacuum, and the potential is given by

\begin{equation}
V=-\frac{A^2\omega^2}{2}.
\label{min}
\end{equation}
On the other hand, the maximum of the potential is given by
\begin{equation}
V=\frac{3A^2}{4}.
\label{max}
\end{equation}
The shape of the potential is shown in Fig.~\ref{po}. In the figure, the horizontal dashed line represents $V=0$. This shows the potential minimum (at $\phi=0$) is an anti-de Sitter vacuum. The red ball with the arrow represents the inflaton field rolling down the potential hill.

Note that the condition $q>\frac{1}{4}$ implies $\omega>0$. In the limiting case $q=\frac{1}{4}$, corresponding to $\omega=0$, the potential in Eq.~(\ref{potential}) with $B=0$ reduces to $V=0$. However, by assuming $\omega \simeq 0$ and expanding the cosine function, the potential can be approximated as
\begin{align}
V(\phi)&\simeq \frac{3}{4}(A^2\omega^2)\phi^2-\frac{1}{2}(A^2\omega^2)   \\
& \equiv \frac{3}{4}\lambda^2\phi^2-\frac{1}{2}\lambda^2. 
\end{align}
This potential coincides with that of uniform rate inflation \cite{Lin:2023xgs}, which corresponds to the solution of the Wheeler-DeWitt equation for $q=\frac{1}{4}$.

The equation of motion is given by \cite{Maki:2026fgm}
\begin{align}
\dot{\phi}=-A\omega \cos (\omega \phi),
\label{phidot}
\end{align}
where we have chosen the initial condition in which a positive $\phi$ rolls toward the minimum at $\phi=0$. This is the origin of the negative sign in the expression for $\dot{\phi}$. This equation is known as the guidance equation in the de Broglie–Bohm interpretation of quantum cosmology. It is possible to integrate the equation of motion and obtain an exact solution for the time evolution of the inflaton field \cite{Maki:2026fgm}. As mentioned in the Introduction, no approximation schemes, such as the slow-roll approximation, are required.
Nevertheless, we will not need to explicitly integrate Eq.~(\ref{phidot}) in the following discussion.

The scale factor $a$ is expressed in terms of a parameter $\alpha$ in the following way.
\begin{equation}
a \equiv e^\alpha.
\label{exa}
\end{equation}
The Hubble parameter is \cite{Maki:2026fgm}
\begin{equation}
H \equiv \frac{\dot{a}}{a}=\dot{\alpha}=\frac{A}{2}\sin(\omega\phi).
\label{adot}
\end{equation}
The last equality is the other guidance equation.
From Eqs.~(\ref{exa}) and (\ref{adot}), we can obtain
\begin{align}
\dot{a}&=e^\alpha \frac{A}{2}\sin(\omega\phi), \\
\ddot{a}&=e^\alpha \left(  \frac{A^2}{4} \sin^2(\omega\phi) -\frac{A^2}{2}\omega^2 \cos^2(\omega\phi)  \right).
\end{align}
By definition, Cosmic Inflation occurs when $\ddot{a}>0$ and $\dot{a}>0$. Inflation ends when $\ddot{a}=0$, which corresponds to $\phi=\phi_e$, and
\begin{equation}
\sec(\omega \phi_e)=\sqrt{2\omega^2+1}.
\label{phi_e}
\end{equation}

\section{Cosmological perturbations}
\label{perturbation}

\begin{figure}[t]
  \centering
\includegraphics[width=0.6\textwidth]{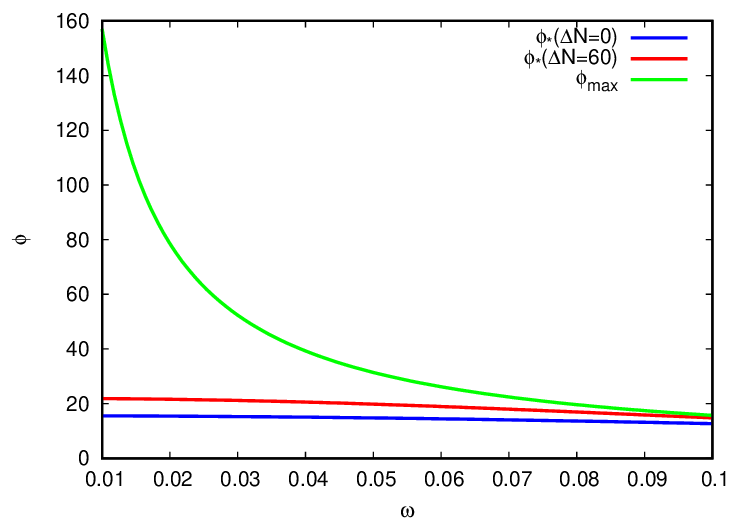}
  \caption{The field value $\phi_\ast$ as a function of $\omega$.}
  \label{con}
\end{figure}

From Eqs.~(\ref{phidot}) and (\ref{adot}), we obtain
\begin{equation}
\frac{d\alpha}{d\phi}=-\frac{1}{2\omega}\tan(\omega \phi).
\label{aphi}
\end{equation}
Physically, $\Delta \alpha=-N$, where $N$ denotes the number of e-folds.
Therefore, 
\begin{align}
\Delta \alpha&=\frac{1}{2\omega}\int^{\phi_i}_{\phi_f}\tan(\omega\phi) d\phi   \\
&=\left.\frac{1}{2\omega^2} \ln | \sec ( \omega\phi )|\right|^{\phi_i}_{\phi_f}.
\end{align}
Here $\phi_i$ denotes the initial value of $\phi$, while $\phi_f$ denotes the field value at the end of inflation. The scale corresponding to the cosmic microwave background (CMB) exits the horizon at $\Delta \alpha \simeq 60$. We define the field value $\phi_*$ at $60$ e-folds before the end of inflation as follows:
\begin{equation}
60=\frac{1}{2\omega^2}\ln\left( \frac{\sec(\omega\phi_*)}{\sec(\omega \phi_f)}\right).
\label{phi_star}
\end{equation}
We are interested in the region where $\cos(\omega \phi)>0$, which implies $\sec(\omega\phi)>0$. Therefore, the absolute value on the right-hand side is unnecessary. Next, we define $K$ for later convenience as follows:
\begin{equation}
K\equiv\sec (\omega \phi_*).
\label{k}
\end{equation}
$K$ depends on $\phi_f$ through Eq.~\eqref{phi_star}. If $\phi_f=\phi_e$, from Eq.~\eqref{phi_e}, we obtain
\begin{equation}
K=\sqrt{2\omega^2+1}e^{2(60)\omega^2}. 
\end{equation}
Alternatively, one may assume that inflation somehow ends at a field value $\phi_0>\phi_e$. For example, assuming that this $\phi_0$ corresponds to the time $40$ e-folds before $\phi_e$, we have
\begin{equation}
40=\frac{1}{2\omega^2}\ln\left(\frac{\sec(\omega\phi_0)}{\sec(\omega \phi_e)}\right),
\end{equation}
which gives $\sec(\omega\phi_0)=\sqrt{2\omega^2+1}e^{2(40)\omega^2}$. Substituting $\phi_f=\phi_0$ into Eq.~\eqref{phi_star}, we obtain
\begin{equation}
K=\sqrt{2\omega^2+1}e^{2(100)\omega^2}.
\end{equation}
More generally, inflation may end $\Delta N$ e-folds before $\phi_e$, rather than fixing $\Delta N=40$. In this case,
\begin{equation}
K=\sqrt{2\omega^2+1}e^{2(60+\Delta N)\omega^2}.
\label{gk}
\end{equation}
A large $\Delta N$ corresponds to a large value of $\phi_\ast$. One may therefore wonder whether $\phi_\ast$ could exceed $\phi_{\mbox{max}}$, where $\phi_{\mbox{max}}$ denotes the field value corresponding to the first maximum of the potential encountered as $\phi$ increases from zero. If $\phi_\ast>\phi_{\mbox{max}}$, the inflaton would roll in the opposite direction. From Eq.~(\ref{potential}) with $B=0$, we find
\begin{equation}
\phi_{\mbox{max}}=\frac{\pi}{2\omega}.
\end{equation}
On the other hand, Eqs.~(\ref{k}) and (\ref{gk}) give
\begin{equation}
\phi_\ast=\frac{1}{\omega}\arccos\left(\frac{1}{\sqrt{2\omega^2+1} e^{2 (60+\Delta N)\omega^2} }\right).
\end{equation}
Since $\arccos \left(\frac{1}{\sqrt{2\omega^2+1} e^{2 (60+\Delta N)\omega^2}} \right)<\pi/2$, we find $\phi_\ast < \phi_{\mathrm{max}}$ for any finite $\Delta N\geq 0$. We plot these field values in Fig.~\ref{con}. It can be seen that the condition $\phi_\ast<\phi_{\mbox{max}}$ is always satisfied throughout the parameter space of interest.

In order to end inflation at $\phi_0$, it may be necessary to introduce an additional mechanism analogous to those employed in hybrid inflation \cite{Linde:1993cn} or hybrid natural inflation \cite{Ross:2016hyb}.
In the following, all observables are evaluated at $\phi_*$. For simplicity, we drop the asterisk and simply denote it as $\phi$.

According to the $\delta N$ formalism \cite{Sasaki:1995aw, Sasaki:1998ug, Lyth:2004gb, Lyth:2005fi, Wands:2000dp}, the primordial curvature perturbation is given by
\begin{equation}
\delta N=-\delta \alpha=-\frac{d\alpha}{d\phi}\delta\phi\mathrel{\simeq}\frac{1}{2\omega}\tan(\omega\phi)\frac{H}{2\pi}=\frac{A}{8\pi \omega}\tan(\omega\phi)\sin(\omega\phi).
\end{equation}
Here, we used $\delta\phi\simeq H/2\pi$ and Eq.~(\ref{adot}) to evaluate the Hubble parameter $H$. The corresponding power spectrum is given by
\begin{align}
P_R&=(\delta\alpha)^2\mathrel{\simeq}\frac{A^2}{64\pi^2\omega^2}\tan^2(\omega\phi)\sin^2(\omega\phi)\\
&=\frac{A^2}{64\pi^2\omega^2}\left(\sec^2(\omega\phi)-1\right)\left(1-\frac{1}{\sec^2(\omega\phi)}\right)\\
&=\frac{A^2}{64\pi^2\omega^2}\left(  -2+K^2+\frac{1}{K^2}  \right) \label{p},
\end{align}
where Eq.~(\ref{k}) has been used in the final equality.

\section{The spectral index and the running spectral index}
\label{index}

\begin{figure}[t]
  \centering
\includegraphics[width=0.6\textwidth]{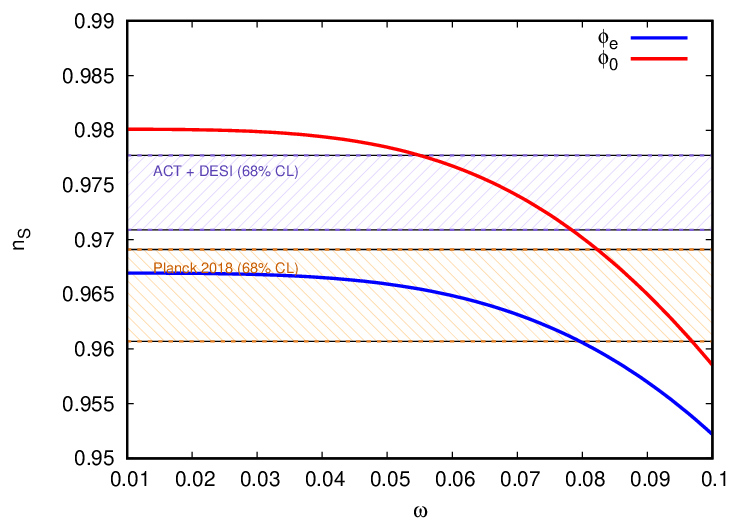}
  \caption{The spectral index as a function of $\omega$. The blue line represents the case where inflation ends at $\phi_e$. The red line represents the case where inflation ends $40$ e-folds before $\phi_e$.}
  \label{id}
\end{figure}

\begin{figure}[t]
  \centering
\includegraphics[width=0.6\textwidth]{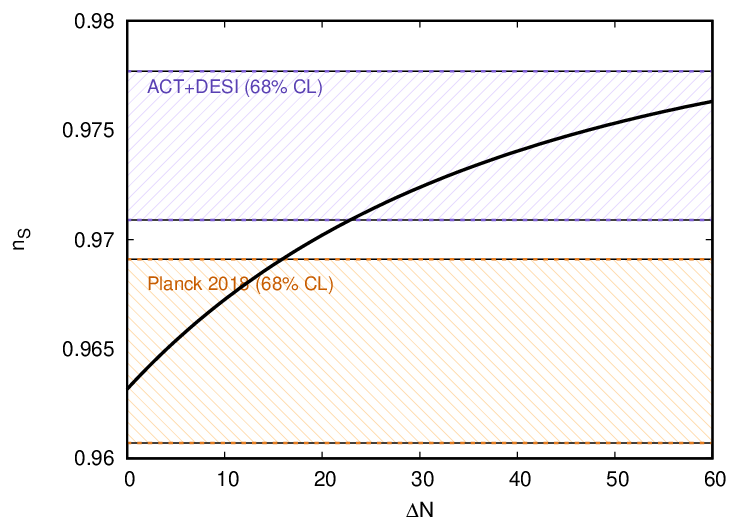}
  \caption{The spectral index as a function of $\Delta N$ for a fixed value of $\omega=0.07$.}
  \label{nsdn}
\end{figure}

The spectral index is given by
\begin{align}
n_S &\equiv 1+\frac{d\ln P_R}{d\ln k} \mathrel{\simeq} 1+\frac{d\ln P_R}{d\alpha}=1+\frac{1}{P_R}\frac{dP_R}{d\phi}\frac{d\phi}{d\alpha}\\
&=1-4\omega^2\left[ \cot^2(\omega\phi)+\csc^2(\omega\phi) \right]\\
&=1-4\omega^2 \left[ \frac{K^2+1}{K^2-1} \right].
\end{align}
In the above derivation, we have used Eqs.~(\ref{aphi}) and (\ref{k}). The spectral index as a function of $\omega$ is shown in Fig.~\ref{id}. 
We also plot the spectral index as a function of $\Delta N$ for a fixed value of $\omega=0.07$, as shown in Fig.~\ref{nsdn}.


\begin{figure}[t]
  \centering
\includegraphics[width=0.6\textwidth]{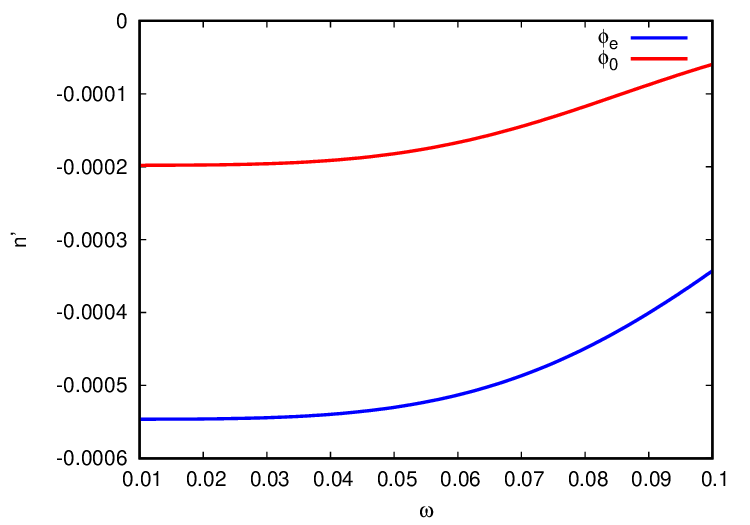}
  \caption{The running spectral index as a function of $\omega$. The blue and red lines correspond to the cases where inflation ends at $\phi_e$ and $40$ e-folds before $\phi_e$, respectively.}
  \label{run}
\end{figure}

\begin{figure}[t]
  \centering
\includegraphics[width=0.6\textwidth]{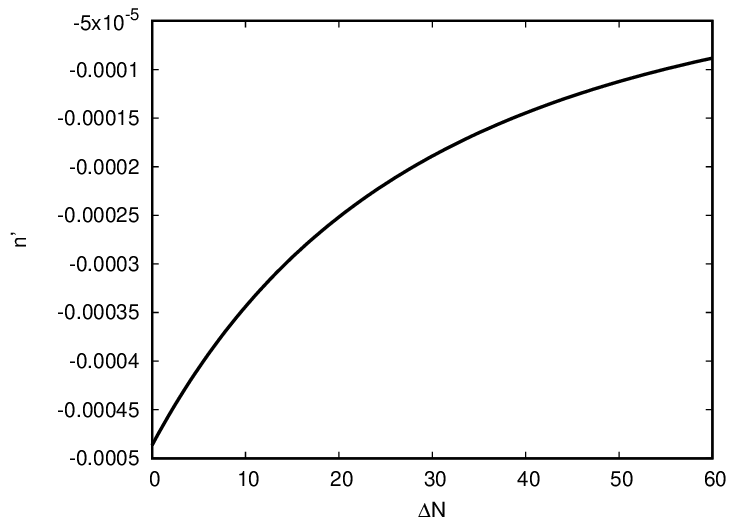}
  \caption{The running spectral index as a function of $\Delta N$ for a fixed value of $\omega=0.07$.}
  \label{dnnp}
\end{figure}

The running spectral index is given by
\begin{align}
n^\prime &\equiv \frac{dn_S}{d\ln k}\mathrel{\simeq}\frac{dn_S}{d\alpha}=\frac{dn_S}{d\phi}\frac{d\phi}{d\alpha}=-32\omega^4 \csc^2(\omega\phi)\cot^2(\omega\phi)\\
&=-32\omega^4\frac{K^2}{\left( K^2-1 \right)^2}.
\end{align}
The running spectral index is plotted in Figs.~\ref{run} and \ref{dnnp}. Observations require $|n^\prime_S| \lesssim 0.01$ \cite{Planck:2018jri}, and our results are well within this bound.


\section{The parameter A}
\label{A}
\begin{figure}[t]
  \centering
\includegraphics[width=0.6\textwidth]{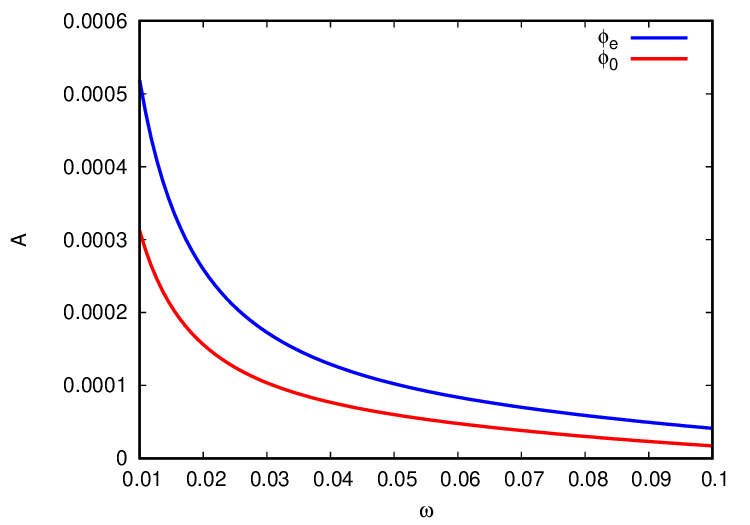}
  \caption{The parameter A as a function of $\omega$. The blue and red lines correspond to the cases where inflation ends at $\phi_e$ and $40$ e-folds before $\phi_e$, respectively.}
  \label{a}
\end{figure}

So far, we have not specified the value of the parameter $A$ appearing in Eq.~(\ref{potential}). It can be determined by imposing CMB normalization $P_R^{1/2}=5\times 10^{-5}$ on the power spectrum. From Eq.~(\ref{p}), we obtain
\begin{equation}
A=\frac{40\pi \times 10^{-5}\omega}{\sqrt{-2+K^2+\frac{1}{K^2}}}.
\end{equation}
This result is plotted in Fig.~\ref{a}. Note that the parameter $A$ determines the maximum of the potential, as shown in Eq.~(\ref{max}). Our results indicate that even the maximum of the potential remains well below the Planck scale.

\begin{figure}[t]
  \centering
\includegraphics[width=0.6\textwidth]{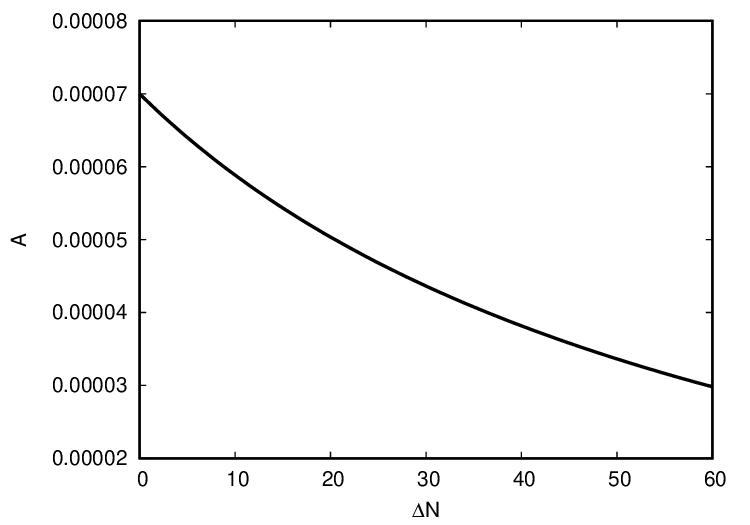}
  \caption{The parameter A as a function of $\Delta N$ for a fixed value of $\omega=0.07$.}
  \label{dna}
\end{figure}

\section{The tensor-to-scalar ratio}
\label{ratio}
\begin{figure}[t]
  \centering
\includegraphics[width=0.6\textwidth]{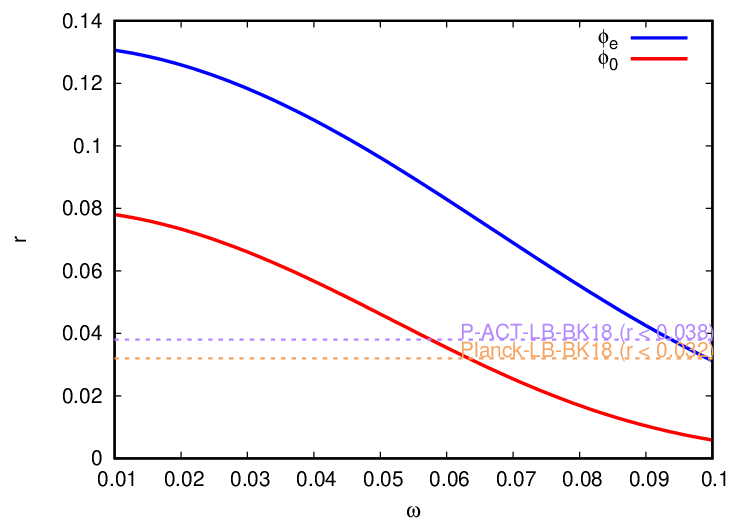}
  \caption{The tensor-to-scalar ratio as a function of $\omega$. The blue and red lines correspond to the cases where inflation ends at $\phi_e$ and $40$ e-folds before $\phi_e$, respectively.}
  \label{r}
\end{figure}

\begin{figure}[t]
  \centering
\includegraphics[width=0.6\textwidth]{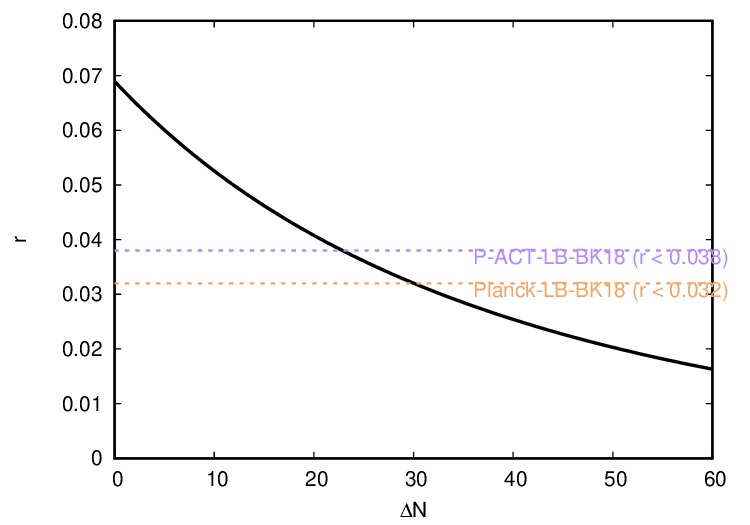}
  \caption{The tensor-to-scalar ratio as a function of $\Delta N$ for a fixed value of $\omega=0.07$.}
  \label{dnr}
\end{figure}

The power spectrum for tensor perturbations is given by 
\begin{equation}
P_T\mathrel{\simeq}8\left( \frac{H}{2\pi} \right)^2=\frac{2H^2}{\pi^2}=\frac{A^2}{2\pi^2}\sin^2(\omega\phi),
\end{equation}
where Eq.~(\ref{adot}) has been used in the final equality.
The tensor-to-scalar ratio is therefore
\begin{equation}
r \equiv \frac{P_T}{P_R}\mathrel{\simeq}\frac{32\omega^2}{K^2-1}.
\end{equation}
The tensor-to-scalar ratio as a function of $\omega$ and $\Delta N$ is shown in Figs.~\ref{r} and \ref{dnr}.


\begin{figure}[t]
  \centering
\includegraphics[width=0.6\textwidth]{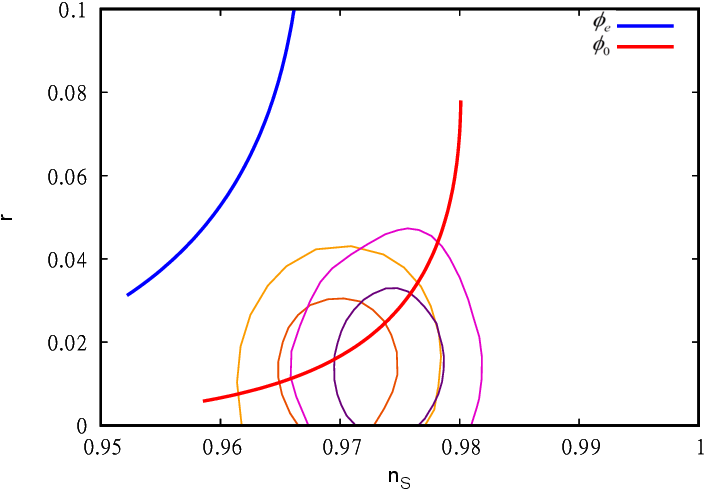}
  \caption{The spectral index and tensor-to-scalar ratio. The parameter $\omega$ is in the range $[0.01,0.1]$. The blue and red lines correspond to the cases where inflation ends at $\phi_e$ and $40$ e-folds before $\phi_e$, respectively.}
  \label{prediction}
\end{figure}

The predictions for both $n_S$ and $r$ are compared with the likelihood contours for Planck-LB-BK18 (orange) and P-ACT-LB-BK18 (purple) from \cite{AtacamaCosmologyTelescope:2025nti} in Fig.~\ref{prediction}. As can be seen from the figure, consistency with observational constraints requires inflation to end at a field value $\phi_0$ before $\phi_e$ is reached. The Simons Observatory is expected to achieve a sensitivity to the tensor-to-scalar ratio of $r < 0.001$ \cite{SimonsObservatory:2025avm}.

The $n_S$ constraint corresponding to the purple contours is shifted upward due to the inclusion of the DESI BAO data. This effect is referred to as the BAO--CMB tension in \cite{Ferreira:2025lrd}, and its resolution remains an open issue.

\section{Conclusion}
\label{Conclusion}
We have proposed a model of a natural inflation with a negative cosmological constant in this work. The potential was not originally introduced for cosmic inflation; rather, it was derived from a solution of the Wheeler-DeWitt equation. A remarkable feature of this model is that the evolution of both the scalar field and the scale factor can be obtained exactly, without relying on approximation schemes such as the slow-roll approximation. 

To achieve consistency with observational constraints, we must treat the field value at the end of inflation, $\phi_0$, as an additional parameter. The allowed range of $\omega$ implies that $q \simeq \frac{1}{4}$. This condition is required to obtain viable predictions for the CMB observables. From Eq.~(\ref{min}), this further indicates that the negative cosmological constant is small compared with the maximum of the potential.

Although this model originated from a study of quantum cosmology, the results obtained here may also be applied to more conventional natural inflation scenarios by introducing a suitably chosen negative cosmological constant.

\appendix

\section{Numerical parameter estimation from CMB observations}
\label{app:ACT}

To complement the analytical study presented in the main text, we perform a numerical Bayesian parameter estimation for our inflation model using the Markov Chain Monte Carlo (MCMC) sampler implemented in the \texttt{Cobaya} package~\cite{Torrado:2020dgo} together with the Boltzmann code \texttt{CAMB}~\cite{Lewis:1999bs}. We consider the same inflationary model discussed in the main text and constrain its parameters using current cosmological observations.

In our analysis, the inflationary observables \(n_s\), \(r\), and \(n^{\prime}\) are not treated as independent sampling parameters. Instead, we sample the model parameters \(\omega\) and \(\Delta N\), from which \(n_s\), \(r\), and \(n^{\prime}\) are calculated at each sampling point according to the predictions of our inflation model. These derived quantities are then passed to CAMB to compute the corresponding scalar and tensor CMB power spectra. Tensor perturbations are included in the analysis.

\begin{figure}[t]
\centering
\includegraphics[width=0.75\textwidth]{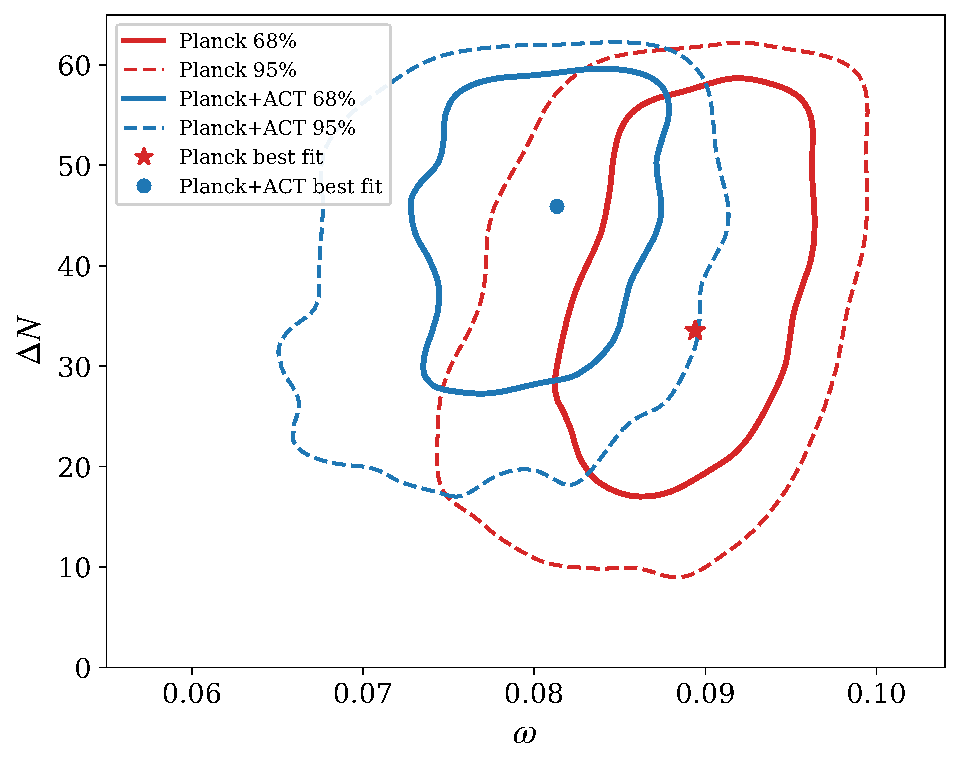}
\caption{
Marginalized 68\% (solid) and 95\% (dashed) confidence regions in the $(\omega,\Delta N)$ plane. The red contours are obtained from the combination of Planck 2018, Planck lensing, DESI BAO, and BK18, while the blue contours additionally include the ACT DR6 likelihood. The corresponding best-fit points are also indicated.
}
\label{fig:omega_DeltaN_ACT}
\end{figure}

We consider two combinations of current cosmological data. The first consists of the Planck 2018 temperature and polarization data, the Planck lensing reconstruction, DESI BAO measurements, and the BK18 $B$-mode polarization data. The second additionally includes the ACT DR6 temperature, polarization, and lensing likelihoods. The resulting marginalized constraints in the $(\omega,\Delta N)$ plane are shown in Fig.~\ref{fig:omega_DeltaN_ACT}.

As expected, the inclusion of the ACT data leads to a moderate shift of the preferred parameter region. In particular, the posterior distribution moves toward slightly smaller values of $\omega$ and correspondingly larger values of $\Delta N$. Since the parameter $q$ is related to $\omega$ through
\begin{equation}
q=\frac14+\frac{\omega^2}{6},
\end{equation}
the ACT data slightly favor values of $q$ closer to $1/4$. Nevertheless, the overall allowed parameter space remains qualitatively unchanged, and the model continues to provide a satisfactory fit to the combined cosmological observations.

For completeness, the posterior means and standard deviations obtained from the two analyses are summarized below:
\begin{align}
\omega &=0.0882 \pm 0.0055,
&
\Delta N &=37.3\pm12.7,
\qquad
({\rm Planck+lensing+DESI+BK18}),
\\
\omega &=0.0795 \pm 0.0056,
&
\Delta N &=42.7\pm10.7,
\qquad
({\rm Planck+ACT+lensing+DESI+BK18}).
\end{align}

Taken together, all the data combinations considered in this work consistently favor the small-$\omega$ region. Although the additional data sets provide slightly tighter constraints, the numerical parameter estimation is in excellent agreement with the physical picture inferred from our previous analytical treatment.

\acknowledgments

This work was in part supported by the National Science and Technology Council (NSTC) of Taiwan under grant number 114-2112-M-167-001 (CML) and JSPS KAKENHI Grants No. JP24K07027 (KK).



\begin{thebibliography}{99}

\bibitem{Starobinsky:1980te}
A.~A.~Starobinsky,
Phys. Lett. B \textbf{91}, 99-102 (1980)
doi:10.1016/0370-2693(80)90670-X

\bibitem{Guth:1980zm}
A.~H.~Guth,
Phys. Rev. D \textbf{23}, 347-356 (1981)
doi:10.1103/PhysRevD.23.347

\bibitem{Linde:1981mu}
A.~D.~Linde,
Phys. Lett. B \textbf{108}, 389-393 (1982)
doi:10.1016/0370-2693(82)91219-9



\bibitem{Maki:2026fgm}
N.~Maki, C.~M.~Lin and K.~Kohri,
[arXiv:2604.25240 [hep-th]].

\bibitem{Lin:2023sza}
C.~M.~Lin,
Chin. J. Phys. \textbf{86}, 344-349 (2023)
doi:10.1016/j.cjph.2023.10.049
[arXiv:2301.06088 [gr-qc]].

\bibitem{Lin:2023xgs}
C.~M.~Lin,
JCAP \textbf{04}, 037 (2023)
doi:10.1088/1475-7516/2023/04/037
[arXiv:2303.04999 [hep-ph]].

\bibitem{Lin:2023ceu}
C.~M.~Lin, R.~Tamura and K.~I.~Nagao,
JCAP \textbf{05}, 105 (2024)
doi:10.1088/1475-7516/2024/05/105
[arXiv:2312.17409 [hep-ph]].

\bibitem{Freese:1990rb}
K.~Freese, J.~A.~Frieman and A.~V.~Olinto,
Phys. Rev. Lett. \textbf{65}, 3233-3236 (1990)
doi:10.1103/PhysRevLett.65.3233

\bibitem{Luu:2025fgw}
H.~N.~Luu, Y.~C.~Qiu and S.~H.~H.~Tye,
Phys. Rev. D \textbf{112}, no.2, 023524 (2025)
doi:10.1103/3mpg-24d2
[arXiv:2503.18120 [hep-ph]].

\bibitem{Luu:2025dax}
H.~N.~Luu, Y.~C.~Qiu and S.~H.~H.~Tye,
JCAP \textbf{09}, 055 (2025)
doi:10.1088/1475-7516/2025/09/055
[arXiv:2506.24011 [hep-ph]].

\bibitem{Shiu:2026edl}
G.~Shiu, F.~Tonioni and H.~V.~Tran,
[arXiv:2604.09141 [astro-ph.CO]].

\bibitem{Murai:2025msx}
K.~Murai and F.~Takahashi,
Phys. Rev. D \textbf{112}, no.10, 103501 (2025)
doi:10.1103/kwhj-vl35
[arXiv:2504.12852 [hep-ph]].

\bibitem{Motohashi:2014ppa}
H.~Motohashi, A.~A.~Starobinsky and J.~Yokoyama,
JCAP \textbf{09}, 018 (2015)
doi:10.1088/1475-7516/2015/09/018
[arXiv:1411.5021 [astro-ph.CO]].

\bibitem{Linde:1993cn}
A.~D.~Linde,
Phys. Rev. D \textbf{49}, 748-754 (1994)
doi:10.1103/PhysRevD.49.748
[arXiv:astro-ph/9307002 [astro-ph]].

\bibitem{Ross:2016hyb}
G.~G.~Ross, G.~German and J.~A.~Vazquez,
JHEP \textbf{05}, 010 (2016)
doi:10.1007/JHEP05(2016)010
[arXiv:1601.03221 [astro-ph.CO]].

\bibitem{Sasaki:1995aw}
M.~Sasaki and E.~D.~Stewart,
Prog. Theor. Phys. \textbf{95}, 71-78 (1996)
doi:10.1143/PTP.95.71
[arXiv:astro-ph/9507001 [astro-ph]].

\bibitem{Sasaki:1998ug}
M.~Sasaki and T.~Tanaka,
Prog. Theor. Phys. \textbf{99}, 763-782 (1998)
doi:10.1143/PTP.99.763
[arXiv:gr-qc/9801017 [gr-qc]].

\bibitem{Lyth:2004gb}
D.~H.~Lyth, K.~A.~Malik and M.~Sasaki,
JCAP \textbf{05}, 004 (2005)
doi:10.1088/1475-7516/2005/05/004
[arXiv:astro-ph/0411220 [astro-ph]].

\bibitem{Lyth:2005fi}
D.~H.~Lyth and Y.~Rodriguez,
Phys. Rev. Lett. \textbf{95}, 121302 (2005)
doi:10.1103/PhysRevLett.95.121302
[arXiv:astro-ph/0504045 [astro-ph]].

\bibitem{Wands:2000dp}
D.~Wands, K.~A.~Malik, D.~H.~Lyth and A.~R.~Liddle,
Phys. Rev. D \textbf{62}, 043527 (2000)
doi:10.1103/PhysRevD.62.043527
[arXiv:astro-ph/0003278 [astro-ph]].

\bibitem{Planck:2018jri}
Y.~Akrami \textit{et al.} [Planck],
Astron. Astrophys. \textbf{641}, A10 (2020)
doi:10.1051/0004-6361/201833887
[arXiv:1807.06211 [astro-ph.CO]].

\bibitem{AtacamaCosmologyTelescope:2025nti}
E.~Calabrese \textit{et al.} [Atacama Cosmology Telescope],
JCAP \textbf{11}, 063 (2025)
doi:10.1088/1475-7516/2025/11/063
[arXiv:2503.14454 [astro-ph.CO]].





\bibitem{SimonsObservatory:2025avm}
I.~Abril-Cabezas \textit{et al.} [Simons Observatory],
JCAP \textbf{04}, 051 (2026)
doi:10.1088/1475-7516/2026/04/051
[arXiv:2512.15833 [astro-ph.CO]].

\bibitem{Ferreira:2025lrd}
E.~G.~M.~Ferreira, E.~McDonough, L.~Balkenhol, R.~Kallosh, L.~Knox and A.~Linde,
Phys. Rev. D \textbf{113}, no.4, 043524 (2026)
doi:10.1103/lq71-b84v
[arXiv:2507.12459 [astro-ph.CO]].

\bibitem{Torrado:2020dgo}
J.~Torrado and A.~Lewis,
JCAP \textbf{05}, 057 (2021)
doi:10.1088/1475-7516/2021/05/057
[arXiv:2005.05290 [astro-ph.IM]].

\bibitem{Lewis:1999bs}
A.~Lewis, A.~Challinor and A.~Lasenby,
Astrophys. J. \textbf{538}, 473-476 (2000)
doi:10.1086/309179
[arXiv:astro-ph/9911177 [astro-ph]].

\end{thebibliography}
\end{document}